\documentclass[fleqn,usenatbib,useAMS]{mnras}

\usepackage{graphicx}	% Including figure files
\usepackage{amsmath}	% Advanced maths commands
\usepackage{amssymb}	% Extra maths symbols
\usepackage{multicol}        % Multi-column entries in tables
\usepackage{bm}		% Bold maths symbols, including upright Greek
\usepackage{pdflscape}	% Landscape pages
\usepackage{caption}

\usepackage{soul}
\usepackage[T1]{fontenc}
\usepackage{ae,aecompl}

\usepackage{newtxtext,newtxmath}
\title[]{Tidal Disruption Events from three-body scatterings and eccentricity pumping in the disks of Active Galactic Nuclei}

\author[Prasad et al.]{Chaitanya Prasad$^1$\thanks{Contact e-mail: \href{mailto:chaitanya.prasad@stonybrook.edu}{chaitanya.prasad@stonybrook.edu}}, Yihan Wang$^{2,5}$, Rosalba Perna$^{1,3}$, K.~E. Saavik Ford$^{3,4,6}$, Barry McKernan$^{3,4,6}$ \\
\vspace{0.1cm}
$^1$Department of Physics and Astronomy, Stony Brook University, Stony Brook, NY 11794-3800, USA\\
$^2$Nevada Center for Astrophysics, University of Nevada, Las Vegas, NV 89154, USA\\
$^{3}$Center for Computational Astrophysics, Flatiron Institute, 
162 5th Ave, New York, NY 10010, USA\\3
$^{4}$Department of Science, BMCC, City University of New York, New York, NY 10007, USA\\
$^{5}$Department of Physics and Astronomy, University of Nevada Las Vegas, Las Vegas, NV 89154, USA\\
$^{6}$Department of Astrophysics, American Museum of Natural History, New York, NY 10024, USA}

\pubyear{2023}

\begin{document}
\label{firstpage}
\pagerange{\pageref{firstpage}--\pageref{lastpage}}
\maketitle

\begin{abstract}
Tidal Disruption Events (TDEs) are routinely observed in quiescent galaxies, as stars from the nuclear star cluster are scattered into 
the loss cone of the central supermassive black hole (SMBH).
TDEs are also expected to occur in Active Galactic Nuclei (AGN),
due to scattering or orbital eccentricity pumping of stars embedded in the innermost regions of the AGN accretion disk. Encounters with embedded stellar-mass black holes (BH) can result in AGN $\mu$TDEs. AGN TDEs and $\mu$TDEs could therefore account for a fraction of observed AGN
variability. Here, by performing scattering experiments with
the few-body code {\tt SpaceHub}, we compute the probability of AGN TDEs and $\mu$TDEs as a result of 3-body interactions between stars and binary BHs. We find that AGN TDEs are more probable during the early life of the AGNs, when rates are $\sim (6\times 10^{-5}-5 \times 10^{-2}) (f_\bullet/0.01)$ $~\rm{AGN}^{-1}$~yr$^{-1}$ (where $f_\bullet$ is the ratio between the number density of BHs and stars), generally higher than in quiescent galactic nuclei. By contrast, $\mu$TDEs should occur throughout the AGN lifetime at a rate of $\sim (1\times 10^{-4} - 4\times 10^{-2})(f_\bullet/0.01)$ $~\rm{AGN}^{-1}$~yr$^{-1}$. Detection and characterization of AGN TDEs and $\mu$ AGN TDEs with future surveys using {\em Rubin} and {\em Roman} will help constrain the populations of stars and compact objects embedded in AGN disks, a key input for the LVK AGN channel.

\end{abstract}


%%%%%%%%%%%%%%%%%%%%%%%%%%%%%%%%%%%%%%%%%%%%%%%%%%

%%%%%%%%%%%%%%%%% BODY OF PAPER %%%%%%%%%%%%%%%%%%

% The MNRAS class isn't designed to include a table of contents, but for this document one is useful.
% I therefore have to do some kludging to make it work without masses of blank space.
%\begingroup
%\let\clearpage\relax
%\tableofcontents
%\endgroup
%\newpage

\section{Introduction}

%Recent progress in wide-field optical and X-ray surveys have discovered rare forms of nuclear variability in galaxies. Otherwise quiescent nuclei can be seen to flare on timescales from days to months, commonly attributed to the disruption and subsequent accretion of material from a close approach star – a so-called tidal disruption event (TDE) \citep{Rees88,Phinney89}. 
Tidal disruption events (TDEs) occur when a star passes close enough to a supermassive black hole (SMBH) to be  disrupted. Around half of the resulting stellar debris self-collides and accretes onto the SMBH \citep{Rees88,Phinney89}. TDEs are widely used to probe distributions of SMBH mass and spin in the local Universe \citep[e.g.][]{vanVelzen2018,Stone2020}, but can also be used to test general relativity \citep[e.g.][]{Ryu2020}, or even light axion models \citep{Du2022}. 
TDEs are routinely discovered in quiescent galactic nuclei (see \citealt{Gezari2021} for a comprehensive observational review). However there is  increased recent interest in their likely presence in Active Galactic Nuclei (AGN).
 
 AGN are powered by the accretion of gas disks onto SMBH. Star formation within the AGN disk adds stars and compact objects directly \citep[e.g.][]{GoodmanTan04,Levin07,Dittmann2020}. Since SMBH are also orbited by nuclear star clusters (NSCs) \citep[e.g.][]{Neumayer20}, a fraction of the NSC must become embedded in the AGN disk, adding to the embedded population over time \citep[e.g.][]{Artymowicz1993,Fabj2020,Nasim23,Wang_2023}. Objects from this embedded population will experience gas torques, migrate, and dynamically encounter each other, leading to collisions, scatterings, and mergers \citep{Bellovary16,Secunda2019,Tagawa2020,Tagawa2021a,Tagawa2021b}. Binary BH mergers in AGN disks should be detected in gravitational waves by LIGO/Virgo \citep[e.g.][]{McKernan2014,Bartos2017,Stone2017,2022Ford}. The LIGO/Virgo AGN channel may account for (among other things): unusual gravitational wave events \citep[e.g.][]{Abbott2020low,Abbott2020high}, the  asymmetric $\chi_{\rm eff}$ distribution \citep{Wang2021chi} (unexpected for a dynamics channel), and the intriguing anti-correlation in the $(q,\chi_{\rm eff})$ distribution \citep{Callister2021, McKernan2022chi}.

AGN disks are therefore expected to contain an embedded stellar population, which may evolve quite differently from stars in vacuo \citep{Cantiello2021,Jermyn2021,Perna2021AIC,Dittman2021}. Complicated dynamical interactions among the embedded populations allow for the possibility of embedded TDEs around the central SMBH \citep[e.g.][]{McKernan2022}, as well as TDEs by stellar-mass BHs within the disk \citep{Yang2022}. TDEs by embedded BH, also known as $\mu$TDEs \citep{Perets2016}, are expected to release energy in a shorter lived but more intense outburst than standard TDEs \citep{Kremer2019,Wang2021TDE}. Certainly the rich (chaotic) dynamics inherently involved in three-body encounters in AGN disks should lead us to expect a wide range of posible outcomes, including TDEs and $\mu$TDEs  \citep{Lopez2019,RyuPap1,RyuPap3,RyuPap4,RyuPap2,Xin2023}. 
AGN are inherently variable, with a large number of potential sources of variability on different time and energy scales. A growing number of AGNs show an extreme degree of variability \citep{Graham17}, flaring or apparently changing their accretion state over timescales incompatible with standard accretion disk models \citep{dexter2018}. Many mechanisms have been proposed to explain extreme AGN flaring, including: microlensing \citep{Hawkins1993}, obscuration changes  \citep{Nenkova2008,Elitzur2012}, nearby supernovae \citep{Kawaguchi1998}, disk instabilities \citep{Penston1984,Shapovalova2010,Elitzur2014}, disk fronts \citep{noda2018,Stern18} or magnetically launched winds \citep{Cannizzaro2020}. TDEs in AGN could provide (yet) another natural explanation \citep{Eracleous1995, Merloni2015, Blanchard2017}, in some fraction of cases. Qualitative estimates of AGN TDE light-curves were initially investigated by \citet{McKernan2022} and predicted signatures may have been observed \citep{Cannizzaro22}. AGN TDEs could therefore be an excellent mechanism to study population dynamics in AGN disks as well as the properties of the disks themselves.

Dynamical interactions may contribute to obtaining TDEs and $\mu$TDEs in AGN disks. Stars scattering off stellar mass BHs in an AGN disk might be sent onto a path towards the SMBH, or be tidally disrupted by the scattering BH itself.  The former scenario is especially favored during the early times of an AGN,  when the stars from the NSC become instantaneously embedded in the disk upon the AGN turns on,
and about half of them will find themselves in retrograde orbits within the disk. Due to drag, these become
highly eccentric, which are then easily driven onto the SMBH \citep{Secunda2021,McKernan2022}.  

While previous dynamical work on the fate of post-scattered stars 
has largely focused on scattering with individual BHs \citep{Wang_2023}, here 
we focus our study on strong dynamical encounters %in the population in the inner disk, particularly 
with BH binaries, given their importance in AGN disks \citep{Samsing2022}. 
This is interesting to study both due to the fact that binaries have a larger geometric cross-section than single stars, as well as because binary formation is enhanced in AGN disks (and hence the star and compact object population is different than that in NSCs).
%Such encounters can lead to TDEs in AGNs with M$_{\rm SMBH}$ $\leq$ 10$^8$ M$_{\odot}$ around non-spinning SMBH. 
By performing detailed scattering experiments 
in various regions of the disk,
we aim to identify regions of parameter space in which BH binary scatterings lead to 
either a TDE by the SMBH, or a $\mu$TDE by the BH binary itself.

The paper is organized as follows:  In Sec.~2 we evaluate whether 
the AGN disk can remain
in a state of full loss cone, to inform us on the proper initial conditions for the scattering experiments. These are described in
Sec.~3, together with our numerical
methods.
Our results are presented in Sec.~4, and we summarize and discuss their astrophysical implications in Sec.~5.

\section{Tidal disruption events around supermassive black holes}
In the following we will discuss the main dynamical processes responsible for loss cone refilling, beginning with the standard case of TDEs from stars in the NSC around the SMBH of a quiescent galaxy (Sec.2.1), and then specializing to the case of AGN disks (Sec.2.2). 
\subsection{TDEs in Nuclear Star Clusters}
As long as orbiting stars in NSCs possess sufficient angular momentum ($L$) they avoid close passes of the SMBH and tidal disruption. TDEs will not occur if $L >L_{\rm min}\sim \sqrt{GM_{\rm SMBH}R_{\rm TDE}}$, where
\begin{equation}
    R_{\rm TDE} = \left(\frac{M_{\rm SMBH}}{M_{\ast}}\right)^{1/3}R_{\ast}
\end{equation}
is the tidal disruption radius of the star of mass $M_{\ast}$ and radius $R_{\ast}$, and
$M_{\rm SMBH}$ is the mass of the SMBH. 
TDEs remove stellar orbits from the 'loss cone' ($L<L_{\rm min}$), while relaxation processes (two-body scattering) within the NSC add stars stochastically to the loss cone.

\subsubsection{Two-body relaxation}
Stars within the radius of influence of the SMBH ($r_{h}$) exhibit approximately Keplerian motion, where \citep{Merritt2013}
\begin{eqnarray}
r_h\sim \frac{GM_{\rm SMBH}}{\sigma^2_{\rm NSC}}=10.75\left(\frac{M_{\rm SMBH}}{10^8M_\odot}\right)\left(\frac{\sigma_{\rm NSC}}{200~{\rm km/s}}\right)^{-2} \rm pc\,,
\end{eqnarray}
and $\sigma_{\rm NSC}$ is the velocity dispersion of the star cluster. Although the SMBH gravitational potential dominates, the potential due to the stars induces stellar orbital precession. The energy and angular momentum of individual stars therefore fluctuate on a relaxation time, $t_{\rm rel}$, given by \citep{Jeans1913,Jeans1916,Chandrasekhar1942,Benney1987}:
\begin{eqnarray}
\frac{\Delta E}{E}\sim \sqrt{\frac{t}{t_{\rm rel}}}\,,\\
\frac{\Delta L}{L_{\rm max}(E)}\sim \sqrt{\frac{t}{t_{\rm rel}}}\,.
\end{eqnarray}
Here, $t_{\rm rel}=\frac{M^2_{\rm SMBH}}{M_*^2N\ln{\Lambda}}T$, where $N$ is the number of stars within the NSC, $\ln\Lambda$ is the Coulomb logarithm, $T$ is the star's orbital period, and
 $L_{\rm max}\sim \sqrt{GM_{\rm SMBH}a}$, with $a$ the semi-major axis of the star's orbit. 

A stellar orbit within the loss cone is consumed by the SMBH during a single orbital period, while the relaxation process introduces fluctuations in angular momentum given by
\begin{eqnarray}
\Delta L_{\rm orb}\sim L_{\rm max}(E)\sqrt{\frac{\log(\Lambda)m^2N}{M^2_{\rm SMBH}}}\,.
\end{eqnarray}
When $\Delta L_{\rm orb} \gg L_{\rm min}$, loss-cone refilling due to relaxation is faster than the consumption of stars in TDEs. This is the 'full loss cone' regime, where the equilibrium TDE rate for a given $E$ is determined by the loss cone size. $L$ is larger on the outskirts of the NSC, so the outer region of the NSC is in the full loss cone regime and this dominates the overall TDE rate in NSCs.
%In the empty loss cone regime (inner NSC)  \citep{Lightman1977, Frank1976, Rees1988, Magorrian1999, Merritt2004,Stone2016},
%\begin{eqnarray}
%\lambda_{\rm empt}(E)\sim \frac{\Delta L_{\rm orb}^2}{L^2_{\rm max}(E)\ln\left(\frac{L_{\rm max}(E)}{L_{\rm min}}\right)}\,,
%\end{eqnarray}

The TDE rate for a given energy $E$ can be expressed using the dimensionless number $\lambda(E)$, which represents the fraction of stars with that particular energy $E$ that undergo disruption per orbital period. In the full loss cone regime, \citep{Lightman1977, Frank1976, Rees1988, Magorrian1999, Merritt2004,Stone2016},
\begin{eqnarray}
\lambda_{\rm full}(E)\sim \frac{\Delta L^2_{\rm min}}{L^2_{\rm max}(E)}\,.
\end{eqnarray}
The overall TDE rate of the NSC is then 
\begin{eqnarray}
\dot{N}_{\rm TDE}\sim\int_{r_c}^{r_h} \frac{\lambda_{\rm full}(E)}{T}n(r) r^2 d\Omega dr\,
\end{eqnarray}
where $r_c$ represents the critical radius where $\Delta L_{\rm orb}=L_{\rm min}$, i.e., the inner boundary of the full loss cone regime, and $N(E)$ is the number of energy states. The overall TDE rate falls within the range of $10^{-4}$ to $10^{-6}$ yr$^{-1}$ for SMBHs with typical masses ranging from $10^{6}$ to $10^{10}$ $M_\odot$, and NSCs with energy profiles that scale approximately as $\frac{dN(E)}{dr}\propto r^{0}$ to $r^{1}$.

Additional effects, such as resonant relaxation \citep{Rauch1996,Rauch1998,Hopman2006,Kocsis2011,Hamers2018} can provide an additional source of TDEs, particularly in nearly empty loss cones. However, in most NSCs this is a sub-dominant effect and we do not consider it further here.

\subsection{TDEs in AGN disks}
In AGN disks, the population of stellar orbits in the energy-angular momentum ($E-L$) phase space is influenced by several processes, and the population of stellar orbits at the loss cone boundary is a crucial factor determining the overall TDE rate.

{ In the following, we adopt an $\alpha$-disk model \citep{Shakura1973}. This can be parameterized in terms of the viscosity parameter $\alpha_d$, the accretion rate efficiency $\lambda_d$, and the Toomre parameter $Q_d$. 
The accretion rate can then be approximated as
\begin{eqnarray}
    \dot{M} = 2.2 \lambda_d \Big(\frac{M_{\rm SMBH}}{10^8 M_{\odot}}\Big) M_{\odot}/yr\,,    \end{eqnarray}
and the surface density follows as
\begin{eqnarray}
\Sigma = \frac{\dot{M}}{2\pi r v_r}\,,
    \label{eq:sigma}
\end{eqnarray}
where
\begin{eqnarray}
v_r = \alpha_d h^2 \sqrt{\frac{GM_{\rm SMBH}}{r}}\,,
    \label{eq:vr}
\end{eqnarray}
with $h\equiv H/r$. 
The scale height $H$ of the disk is given by
\begin{eqnarray}\label{eq:H}
     H = \Big(\frac{Q_d \dot{M}}{2 \alpha_d M_{\rm SMBH} \Omega_d}\Big)^{1/3} r\,,
\end{eqnarray}
where r is the distance from the SMBH, and $\Omega_d=v_r/r$ is the orbital frequency. 
}

\subsubsection{Eccentricity Damping due to Co-orbital gas}
Prograde stellar orbits in AGN disks experience torques from co-orbiting gas, which causes damping of their orbital eccentricity. A similar process acts on planets embedded in proto-planetary disks \citep{Goldreich03}. The damping timescale 
 is \citep{Tanaka2004,Arzamasskiy2018}
\begin{eqnarray}
t_{\rm e,damp}&\sim&h^2\frac{M_{\rm SMBH}}{\Sigma \pi r^2}\frac{M_{\rm SMBH}}{M_*}T= Q_dh\frac{M_{\rm SMBH}}{M_*}T\nonumber\\
&=& 0.42Q_d^{4/3}\left(\frac{\lambda_d}{\alpha_d}\right)^{1/3}\left(\frac{r}{1000\rm au}\right)^{2}\left(\frac{M_*}{1M_\odot}\right)^{-1}\nonumber\\
&&\left(\frac{M_{\rm SMBH}}{10^8M_\odot}\right)^{1/3}\rm Myr\,.
\label{eq:lindblad}
\end{eqnarray}
 Due to eccentricity damping, fully embedded {\em{prograde}} orbits tend to reach their maximum angular momentum for a given orbital energy, which pushes the orbits away from the loss cone.

\subsubsection{Refilling of Disk Stars from the NSC}
In addition to the initially fully embedded prograde orbits, the interaction between the disk and the NSC results in the capture of stars from the NSC into the AGN disk. These captured stars refill the $E-L$ phase space, providing extra relaxation for orbits within the AGN disk. The capture process is characterized by a typical timescale \citep{Generozov2023,Wang2023b}
\begin{eqnarray}
t_{\rm cap}\sim \frac{\Sigma_*}{\Sigma}T=&&0.7Q_d^{2/3}\left(\frac{\alpha_d}{\lambda_d}\right)^{1/3}\left(\frac{M_*}{1M_\odot}\right)\left(\frac{M_{\rm SMBH}}{10^8M_\odot}\right)^{-4/3}\nonumber\\
&&\left(\frac{R_*}{R_\odot}\right)^{-2}\left(\frac{r}{1000\rm au}\right)^{3}\rm Myr\,,
\end{eqnarray}
where $\Sigma_{\ast}={M_{\ast}}/{\pi R_{\ast}^2}$ is the surface density of stars of  radius $R_*$. The stellar refilling process from capture can be described as \citep{Wang2023b}:
\begin{eqnarray}
M_{\rm cap} &\sim& 2M_{\rm SMBH}f_*\left(\frac{\Sigma}{\Sigma_*}\frac{t}{T_h}\right)^{1-\gamma_{\rm NSC}/3}\\
\frac{dN}{dr}&\propto& r^{-1/4}\,.
\end{eqnarray}
Here, $M_{\rm cap}$ is the total captured stellar mass in the AGN disk, $f_*$ is the mass fraction of stars in the NSC, 
%$\Sigma_h$ is the surface density of the AGN disk at $r_h$, 
$T_h$ is the orbital period at $r_h$, and $\gamma_{\rm NSC}$ is the radial number density profile power-law index of the NSC.

During the disk-star interactions,  eccentricity damping drives captured stars (overwhelmingly prograde orbits) into nearly circular orbits. Thus, most  captured stars populate the region near $L_{\rm max}$, which is far away from the loss cone. So the disk capture process is unable to efficiently refill the loss cone.

\subsubsection{Two-body relaxation from in-disk single-single scatterings}
Two-body relaxation (via close encounters and remote perturbations) could also relax the orbital angular momentum of disk stars away from $L_{\rm max}(E)$, potentially generating TDEs. However, the total number of stars in an AGN disk is a small fraction of the total number of stars in an NSC, yielding only small fluctuations in the $E-L$ phase space. The two-body relaxation timescale in an AGN disk is approximately given by
\begin{eqnarray}
t_{\rm rel,disk} &\sim& \frac{v^2_{\rm orb}(a)}{\dot{v}^2_{\rm orb}(a)}\sim \frac{a^3}{\bar{b}(a)^3}T\nonumber \\
&=& 776 Q^{-2/3}\left(\frac{\lambda_d}{\alpha_d}\right)^{1/3}\left(\frac{M_*}{1M_\odot}\right)^{-1}\left(\frac{M_{\rm SMBH}}{10^8M_\odot}\right)^{10/3}\nonumber\\
&&\left(\frac{R_*}{R_\odot}\right)^{2}\rm Gyr\,, \label{eq:dv}
\end{eqnarray}
where $\bar{b}(a)$ is the average impact parameter for scattering between two prograde circular orbits around semi-major axis $a$. Since $t_{\rm rel,disk} \gg t_{\rm e,damp}$ in eqn.~\ref{eq:lindblad}, orbital eccentricity damping is much faster than the relaxation process. So in-disk two-body relaxation cannot efficiently fill the AGN loss cone with stellar orbits.

\subsubsection{TDE from retrograde orbits}
Approximately half of the initial orbits in a newly formed AGN disk will be retrograde with respect to the flow of the disk gas \citep{McKernan2022}. Stars on these retrograde orbits will experience strong aerodynamic drag. In low density regions of the AGN disk, both the eccentricity and the semi-major axis of the star decrease. However, in specific AGN disk models with high density in the inner regions (e.g., the classic Shakura-Sunyaev disk, \citealt{Shakura1973}), the timescale for semi-major axis decay can become shorter than the orbital period. In such cases, the eccentricity would increase \citep{wang2023stellarbh}, akin to a rapid inspiral. As these decaying retrograde orbits shrink to the tidal disruption radius, tidal disruption events can be triggered. Once these retrograde orbits are consumed by TDEs, they cannot be efficiently replenished through the disk capture process, as the objects captured in AGN disks are in prograde orbits. Therefore, this burst of retrograde TDEs can occur only in the early stages after the AGN turns on.
 
The total budget of initial retrograde orbits available for TDEs, given a disk  scale height $h(r)$, is
\begin{eqnarray}
M_{\rm ret}\sim \int_{r_{\rm TDE}}^{r_{\rm disk,out}} dr \int_{-1}^{-1+h^2(r)/2}d\cos I 2\pi r^2 \rho_{\rm NSC}(r)\,,
\end{eqnarray}
where $\rho_{\rm NSC}$ is the density of the NSC. The total mass budget for the retrograde orbiting stars is $\sim {(2/3)h(r_{h})}M_{\rm SMBH}$ for a NSC cluster with radial density power-law index $\gamma_{\rm NSC}=2$.
The timescale for orbital decay of the  retrograde stars is
\begin{eqnarray}
t_{\rm ret}\sim\frac{\Sigma_*}{\Sigma}hT&=&10^{-3}\left(\frac{M_*}{1M_\odot}\right)\left(\frac{M_{\rm SMBH}}{10^8M_\odot}\right)^{-3/2}\left(\frac{R_*}{R_\odot}\right)^{-2}\nonumber\\
&&\left(\frac{r}{1000\rm au}\right)^{7/2}\rm Myr\,.
\end{eqnarray}
Since $t_{\rm ret}$ is very short  compared to the expected lifetime of the AGN (few Myrs), we should expect an increased TDE rate very early in the AGN lifetime, before the budget for retrograde orbits has been depleted. The rate of these TDEs will be computed in Sec.~3.

\subsection{Micro-TDEs}\label{sec:utde}
Apart from the normal TDEs where stars are tidally disrupted by SMBHs, within the AGN disk the high density of the stellar population makes the probability of a star being tidally disrupted by a stellar-mass black hole significant.

The two-dimensional (2D) cross-section of a $\mu$TDE in a specific environment can be estimated by the following equation:
\begin{eqnarray}
\sigma_{\mu \rm TDE}\sim \pi R_{\rm \mu TDE}^2\left(1+\frac{r_\sigma}{R_{\rm \mu TDE}}\right)\,, 
\end{eqnarray}
where $R_{\rm \mu TDE}=(\frac{3M_\bullet}{M_*})^{1/3}R_*$ is the tidal radius of the star within the influence of the stellar BH  of mass $M_\bullet$, and $r_\sigma \sim\frac{ 2G(M_\bullet+M_*)}{\sigma^2_{\rm env}}$ represents the two-body interaction radius in the environment, where gravitational forces between two objects dominate over the background potential created by other objects, with velocity dispersion denoted by $\sigma_{\rm env}$. The ratio {$r_\sigma/R_{\rm \mu TDE}$} measures the strength of the gravitational focusing effect.

For objects in the NSC, the velocity dispersion scales as $\sim 2.3(M_{\rm SMBH/}M_{    \odot})^{1./4.38}$~km/s 
\citep{Kormendy2013}, which results in $r_\sigma\sim 1000R_\odot$ for a $10^8M_\odot$ SMBH, $60M_\odot$ stellar mass BH and $1M_\odot$ main sequence star. Consequently, the gravitational focusing factor for objects in the NSC is {$1+r_\sigma/R_{\rm \mu TDE}\sim 200$}.

As stars and BHs are captured by the AGN disk, they will approximately align within the same orbital plane, making the cross-section one-dimensional (1D), and thus
the corresponding (1D) focusing factor to be proportional to {$\sqrt{1+r_\sigma/R_{\rm \mu TDE}}$}. 
For scatterings between prograde orbits within the AGN disk, the velocity dispersion will be reduced by a factor of $\bar{b}/r$ compared to the velocity dispersion in the NSC, {leading to a larger $r_\sigma$ by a factor of $r/\bar{b}$. } For scatterings between one prograde orbit with another retrograde orbit (retrograde scatterings), the velocity dispersion will  be $\sim$ 2 times the local Keplerian velocity; as a result, the 1D $\mu$TDE cross-section for retrograde scatterings within the AGN disk will be a few ($\sim\sqrt{1+200/4}$) of the 1D geometric $\mu$TDE cross-section $R_{\rm \mu TDE}$. { The order of magnitude estimate provided here will be numerically quantified in Sec.\ref{sec:results}.}

\section{Scattering Experiments}

\begin{figure*}
    \includegraphics[width=\textwidth]{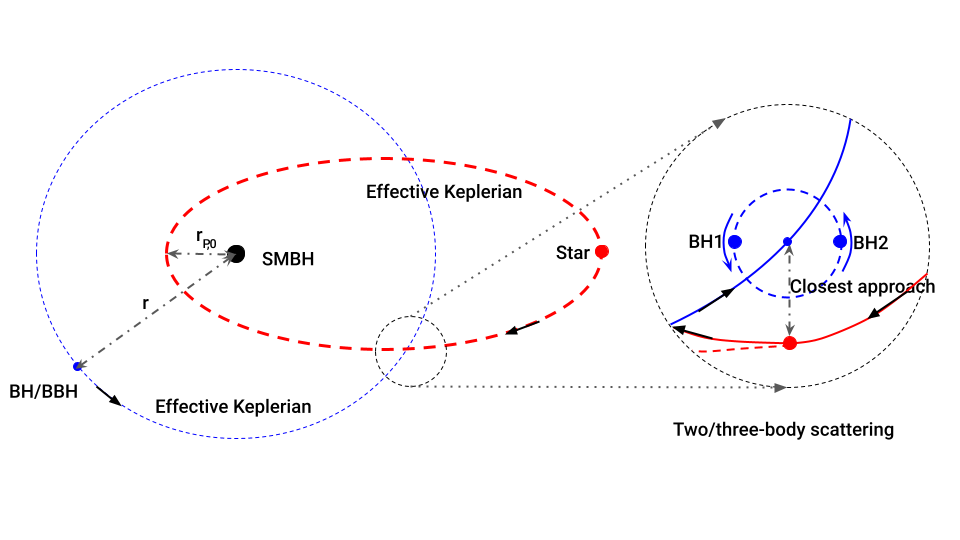}
    \caption{Schematics of the two/three-body scattering experiments in the central potential of a SMBH. \textit{Left:} panel shows the BH/BBH in a prograde circular Keplerian orbit around the SMBH at distance $r$, and the star in a retrograde elliptical Keplerian orbit with the same semi-major axis $r$ and pericenter distance $r_{P,0}$. The dashed lines represent orbits when no interaction takes place. \textit{Right:} panel shows the two/three-body interaction between the BH/BBH and the star with a closest distance of approach between the star and BH/center of mass of the BBH, zooming in at the scattering location, resulting in the star's orbit getting altered. The dashed red line represents the orbit of the star if no interaction takes place, and the solid red line represents the altered orbit.}
    \label{fig:schematic}
\end{figure*}

Since we are focusing here on estimating the rate of TDEs and $\mu$TDEs in an AGN disk, our numerical experiments are performed in a coplanar geometry. The general set up is that of a central SMBH, a binary BH  scatterer, and an incoming tertiary star, as schematically shown in Fig.~\ref{fig:schematic} and further detailed in Sec.~3.2.
To better appreciate the effect of binarity on the TDE and $\mu$TDE cross-sections, we also perform comparative experiments in which the scatterer is a single BH of mass equal to the sum of the masses of the two BHs when in a binary (detailed in Sec.3.1). Our experiments are focused on encounters with retrograde stars embedded in the disk shortly after the AGN turns on, since these are by far the most probable source of AGN TDEs.

The scattering experiments are performed using the high-precision, few-body code {\tt SpaceHub} \citep{2021Spacehub}. For this work, since the mass ratio between the BHs and the SMBH is very small, and the eccentricities are large, we use the AR-chain$^{+}$ method of integration \citep{Mikkola1993}. We ignore the post-Newtonian corrections, since the distances between the objects are large enough, and the results from the simulation show no significant differences if we include these corrections.

Note that our scattering experiments are purely gravitational and neglect hydrodynamical interactions (e.g. Lindblad resonance, corotation resonance, dynamical friction and aerodynamic drag) on the timescales of the scatterings. The Lindblad, corotation resonances, dynamical friction and aerodynamical drag act on a typical timescale that is longer than the orbital period, while the scattering timescale is typically shorter than the orbital period. Therefore, they can be safely ignored during the scattering processes. However, we note that this assumption does not apply to resonance scatterings (i.e., multiple scatterings), where the scattering can last longer than the orbital period around the SMBH, making those dissipation processes important. For example, \citet{Li2023} and \citet{Rowan2023} have shown that dynamical friction can facilitate the formation of binary BHs in AGN disks if the two interacting objects undergo resonance scattering.

\subsection{Single-single scattering}
We set up coplanar scattering experiments between a BH and a star, both orbiting a SMBH, whose mass is $M_{\rm SMBH}=10^{8}M_{\odot}$. 
This choice is motivated by the fact that most AGN masses are in the 
$\sim 10^7-10^8 M_\odot$  range
\footnote{http://www.astro.gsu.edu/AGNmass/}. We reserve a discussion of the dependence on the SMBH mass to Sec.~4.3. The BH has a mass $M_{\rm BH}=60M_{\odot}$ and lies in a circular orbit ($e_{\rm BH}=0$) around the SMBH.

The star is initially in a highly eccentric elliptical orbit of semi-major axis equal to the orbital distance of the BH\footnote{For retrograde encounters, which, as we will show, are dominated by highly eccentric orbits, the bulk of the TDE events comes from the relatively more nearby stars to the scatterer, which are captured by this initial condition. }, and has a mass $M_{\ast}=M_{\odot}$ and radius  $R_{\ast}=R_{\odot}$. The initial pericenter distance of the star from the SMBH is increased (by decreasing the eccentricity) starting from the minimum value of $r_{P,0}=$ 1$R_{\rm TDE}$. If the post-scattering pericenter distance of the star from the SMBH falls within this tidal disruption radius $R_{\rm TDE}$, a TDE is triggered. Similarly a $\mu$TDE is triggered if the distance between the star and the BH is smaller than $R_{\mu{\rm TDE}}$ at any time. 
We repeat the scattering experiments by placing the BBH across a range of disk distances: $r=[10^{2},10^{5}]r_{g}$. {This range of radii is well within what expected from detailed modeling of AGN disks \citep{Sirko2003,Thompson2005}}.

\subsection{Binary-single scattering}
For binary-single scattering, we take the same setup and replace a single BH with a binary BH with equal total mass ($60M_{\odot}$). The BBH is in a perfectly circular ($e_{\rm BBH}=0$) prograde orbit. We explore a range of BBH radial locations from $r=10^2 r_g$ to $10^5 r_g$ and three different sized binaries $a_{\rm BBH}=[0.01,0.1,1]R_{\rm H}$ where
\begin{eqnarray}
    R_{\rm H} = \left(\frac{M_{\rm BBH}+M_{*}}{3M_{\rm SMBH}}\right)^{1/3}r
\end{eqnarray}
is the mutual Hill radius of the BBH and the star.

Similarly, the initial semi-major axis of the orbit of the star was set up to be the same as the orbit of the BBH around the SMBH. The initial pericenter distance of the star from the SMBH is varied from $r_{P,0}=$ 1$R_{\rm TDE}$ to 5~$R_{\rm TDE}$, resulting in a highly eccentric retrograde orbit. For each set of parameters, one million scattering experiments were performed to obtain statistically significant results.

\section{Results}\label{sec:results}

\subsection{Scattering outcomes via phase-space diagrams}
The scattering experiments were performed by fixing the distance $r$ between the SMBH and the BHs, and the initial pericenter distance of the star $r_{P,0}$. This effectively fixes the initial eccentricity of the orbit, since $e = 1-r_{P,0}/r$ for an elliptical orbit. For single-single scattering, the only variable parameter then is the closest distance of approach between the star and BH, or equivalently the impact parameter. 

For binary-single scattering, the initial phase $\phi$ of the BBH introduces a new variable to the problem (in addition to the distance of closest approach between the star and the BBH). We now vary the distance of closest approach to the center of mass of the binary (rather than the distance to a single BH as in the single-single scatterings), and $\phi$, to explore areas of parameter space where TDEs or $\mu$TDEs could be triggered.

\begin{figure*}
    \includegraphics[width=\columnwidth]{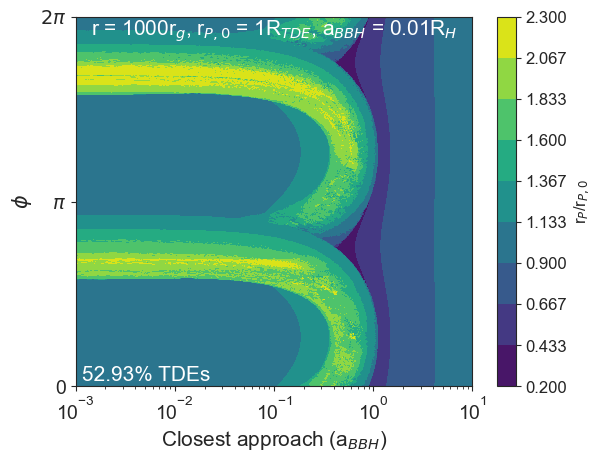}\includegraphics[width=\columnwidth]{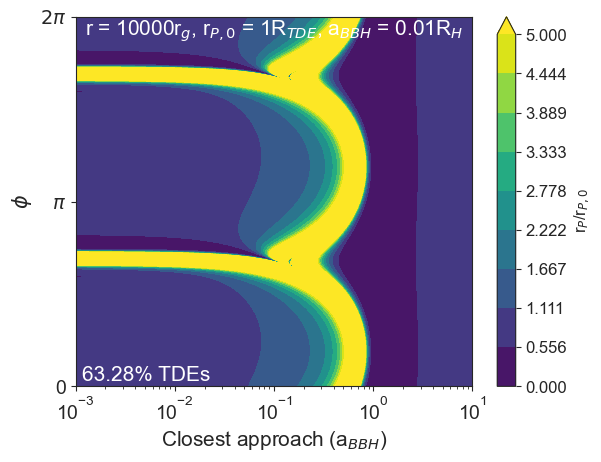}
    \includegraphics[width=\columnwidth]{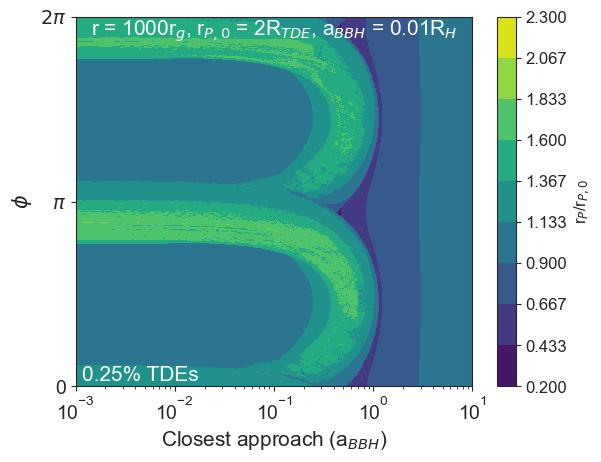}\includegraphics[width=\columnwidth]{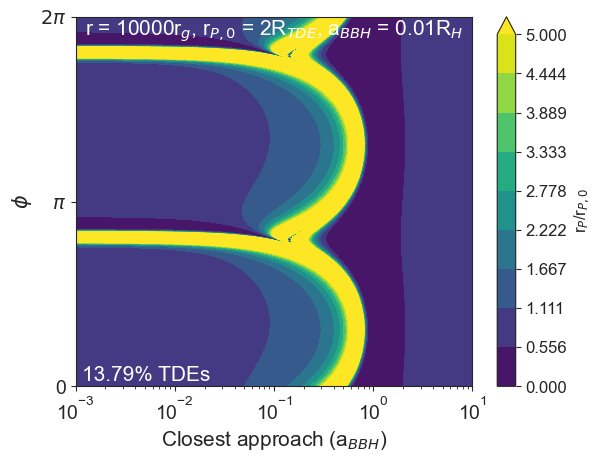}
    \caption{Phase space diagrams illustrating the    
    results of the scattering experiments between a star and a BBH of size $a_{\rm BBH}=$ 0.01$R_{\rm H}$. The colorbar indicates the ratio between the final pericenter distance and the initial pericenter distance of the star from the SMBH, as a function of the closest approach distance between the star and the center of mass of the BBH in units of the BBH size and the initial phase $\phi$ of the BBH. Above each figure is listed the scattering location $r$, the initial pericenter distance of the star from the SMBH $r_{P,0}$, and the size of the BBH $a_{\rm BBH}$. At the bottom of each figure is listed the percentage of the phase space in which a TDE is observed.}
    \label{fig:BBH3}
\end{figure*}

Figure \ref{fig:BBH3} shows our results for the smallest BBH size $a_{\rm BBH}=$ 0.01$R_{\rm H}$. The left column shows our results for the distance $r$ = $10^3r_{g}$, the right column shows the same for $r$ = $10^4r_{g}$. The colorbar indicates the ratio of the final pericenter distance to the initial pericenter distance of the star from the SMBH. At the bottom of each plot, we have indicated the percentage of scenarios that result in TDEs, {up to the given maximum closest approach, which is equal to 10$a_{\rm BBH}$ in the experiments}. The top row shows cases with initial pericenter distance $r_{P,0}=1R_{\rm TDE}$, and the bottom row shows cases with $r_{P,0}=2R_{\rm TDE}$.

In Figure \ref{fig:BBH3}, we can see that as we increase $r_{P,0}$ (from top to bottom), the parameter space resulting in TDEs decreases sharply. This is not surprising since the initial orbit of the star has a pericenter that is farther from the SMBH, so the interaction with the BBH is rarely strong enough to scatter the star to within the tidal disruption radius of the SMBH. Indeed, when we further increase the initial pericenter distance to $r_{P,0}=5R_{\rm TDE}$, no TDEs are observed, and thus the results have not been included.

\begin{figure*}
    \includegraphics[width=\columnwidth]{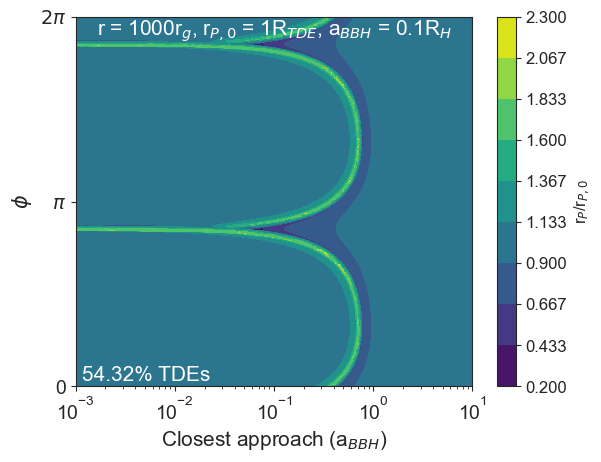}\includegraphics[width=\columnwidth]{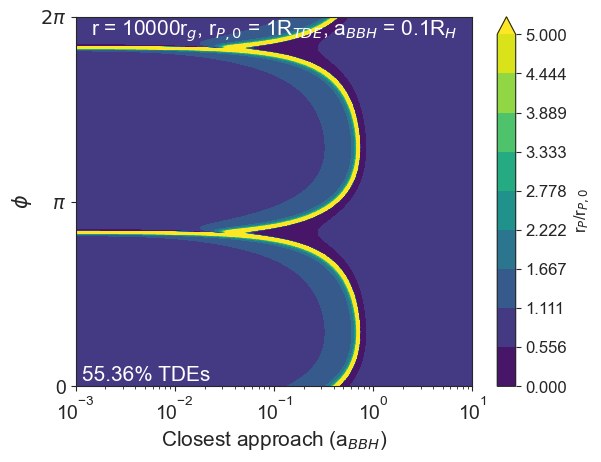}
    \includegraphics[width=\columnwidth]{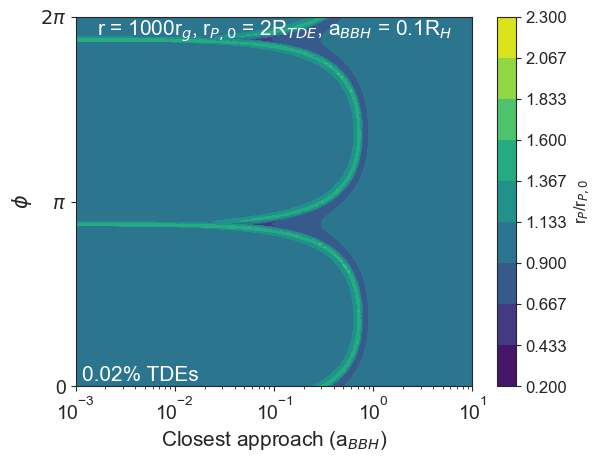}\includegraphics[width=\columnwidth]{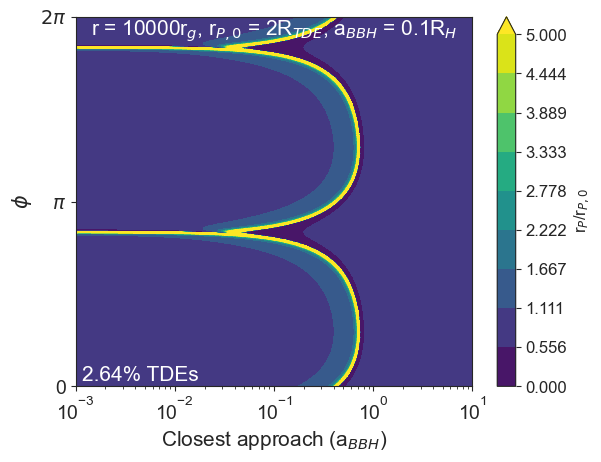}
    \caption{Same as Figure \ref{fig:BBH3} but with BBH of size $a_{\rm BBH}=$ 0.1$R_{\rm H}$.}    
    \label{fig:BBH2}
\end{figure*}

Similarly, Figure \ref{fig:BBH2} shows our results for the BBH size $a_{\rm BBH}=$ 0.1$R_{\rm H}$, while the rest of the parameters are the same as in Figure \ref{fig:BBH3}. We see a similar trend with $r_{P,0}$ regardless of the size of the binary, though this larger binary generates fewer overall TDEs. However, when the size of the BBH is smaller,  {the BBH has steeper gravitational potential} and thus scatterings can be stronger. This can be seen by comparing the results in Figures \ref{fig:BBH3} and \ref{fig:BBH2}. Comparing the same set of initial parameters in the two figures, the fraction of scenarios resulting in TDEs, i.e. the star getting scattered closer to the SMBH, increases for smaller $a_{\rm BBH}$. The fraction of scenarios where the star gets scattered farther away from the SMBH also increases, which is evident from the maximum value of the colorbar.

A third scenario was explored, where the BBH size $a_{\rm BBH}=$ 1.0$R_{\rm H}$, but it has not been included in the results since this BBH is easily disrupted by the incoming star, and does not produce any significant number of TDEs. 

All the scenarios shown in Figures \ref{fig:BBH3} and \ref{fig:BBH2} have stars with highly eccentric pre-scattering orbits ($e \approx 1$), going almost directly towards the SMBH, with difference in eccentricites of the order $10^{-2}$. However, the number of TDEs sharply drops going from the top row to the bottom row as the initial eccentricity of the star is decreased minutely. This shows that TDEs resulting from scattering dynamics require highly initial eccentricity, almost radial initial orbits. Our results show that the change in the orbit of the star as a result of scattering isn't significant enough in most cases, as the ratio of the final and the initial pericenter distance is of order of $\sim 1$
$\sim 10^{0}$ 
in most scenarios.

\begin{figure*}
\includegraphics[width=\columnwidth]{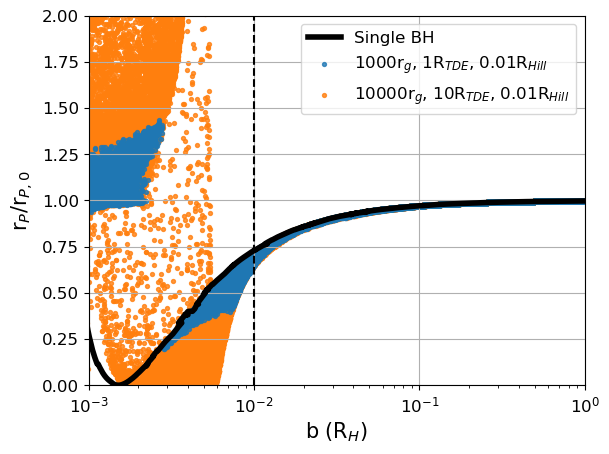}\includegraphics[width=\columnwidth]{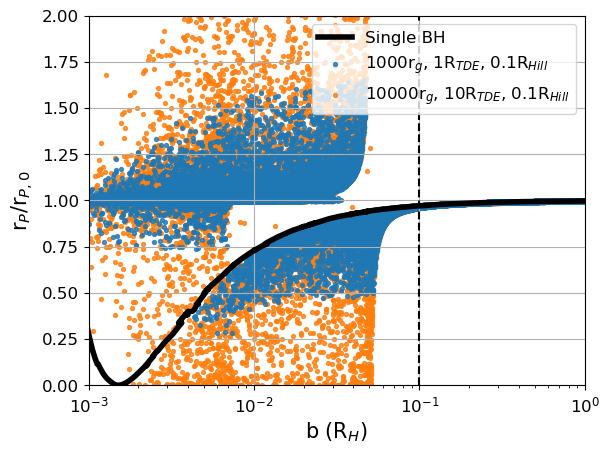}
    \caption{Ratio of the final pericenter distance of star from SMBH to the initial pericenter distance as a function of the impact parameter $b$. The solid line shows results for a single BH scatterer at $r=10^3r_g$ and the initial pericenter distance of the star $r_{P,0}=1R_{\rm TDE}$. The left panel shows the results for BBH of size $a_{\rm BBH}=$ 0.01$R_{\rm H}$, while the right panel shows the same for BBH of size $a_{\rm BBH}=$ 0.1$R_{\rm H}$, the dashed vertical lines mark the size of the BBH. The two sets of parameters shown in both the figures are: [$r=10^3r_g$, $r_{P,0}=1R_{\rm TDE}$] and [$r=10^4r_g$, $r_{P,0}=10R_{\rm TDE}$].}
    \label{single-single}
\end{figure*}

We now compare our results from the binary-single scattering to those from the single-single scattering.  Since our interest is in the production of TDEs, we restrict ourselves to the parameter space for which scatterings in the single BH case lead to the star being scattered towards the SMBH rather than away from it. This gives a conservative cross-section within a factor of $\lesssim 2$ for the BBH scatterer, and no difference for the single BH one.
In Figure \ref{single-single}, we show the ratio of the final pericenter distance and the initial pericenter distance of the star from the SMBH. The black solid line shows our results for a single 60$M_{\odot}$ BH scattering a star at a distance $r$ = $10^3 r_g$ and initial pericenter distance $r_{P,0}$ = 1$R_{\rm TDE}$, corresponding to an initial eccentricity $e = 1-r_{P,0}/r \approx 0.99$. We compare this with our results for a 30$M_{\odot}$ - 30$M_{\odot}$ BBH scattering a star with the same initial eccentricity as the single BH scenario. {For the binary scenario, we adopted two different binary sizes, 0.01 $R_{\rm H}$ and 0.1 $R_{\rm H}$.} The left plot shows results for the BBH size $a_{\rm BBH}=$ 0.01$R_{\rm H}$, and the right plot shows results for $a_{\rm BBH}=$ 0.1$R_{\rm H}$, and vertical black dashed line in each plot indicates the size of the BBH. In each plot, the blue dots show results from BBH scatterings with the same initial parameters as those in the single BH scattering experiment: $r$ = $10^3r_{g}$, $r_{P,0}$ = $1R_{\rm TDE}$, while the orange dots show results from BBH scatterings with: $r$ = $10^4r_{g}$, $r_{P,0}$ = $10R_{\rm TDE}$, thus effectively keeping the initial eccentricity of the star fixed.

\begin{figure*}
    \includegraphics[width=\columnwidth]{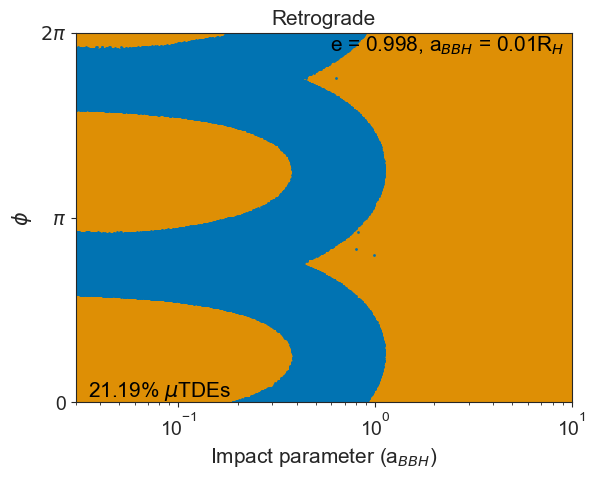}\includegraphics[width=\columnwidth]{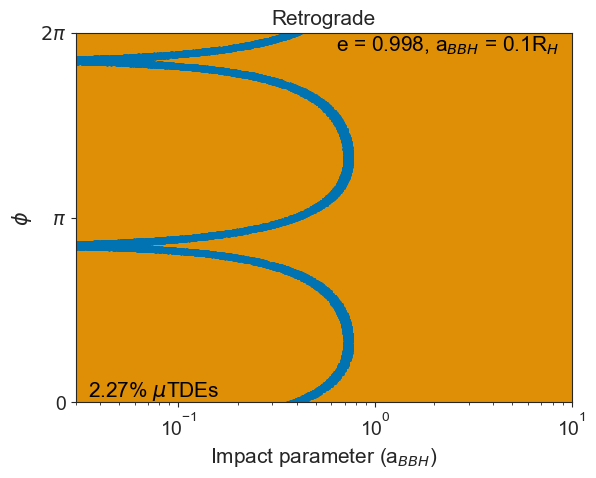}
    \includegraphics[width=\columnwidth]{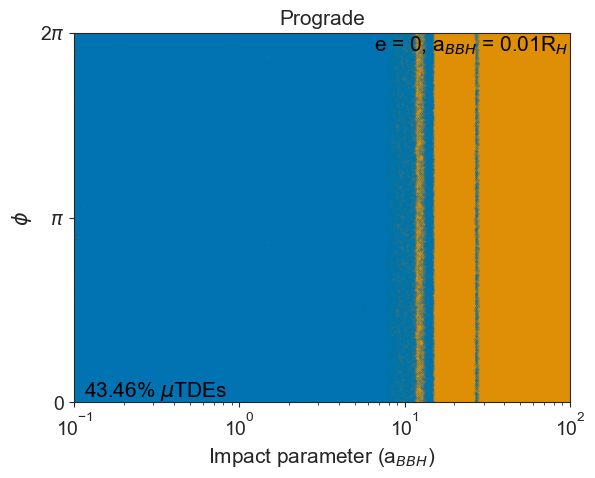}\includegraphics[width=\columnwidth]{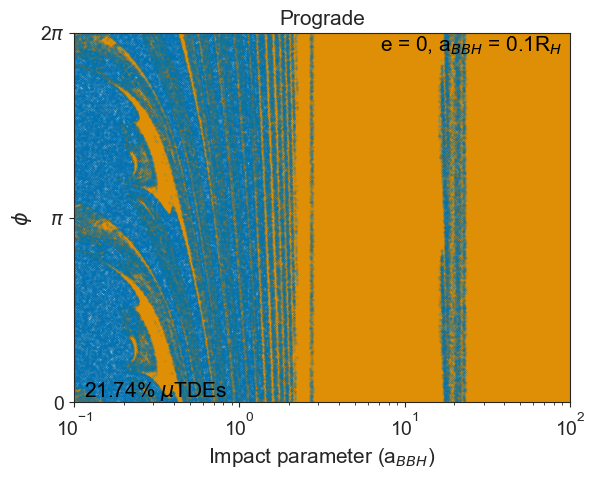}
    \caption{Phase diagrams illustrating the regions of initial binary phases and impact parameters for which scatterings lead to a $\mu$TDE (blue dots) rather than to a simple scattering (orange dots). 
    The BBH scatterer is placed at a radial distance $r=10^3r_g$. Stars in prograde orbital motion are assumed to have an initial eccentricity $e=0$, while stars in retrograde motion come in nearly radial orbits.  The cases shown here have initial eccentricity $e=0.998$ as a representative example. Left and right panels illustrate how the scattering outcome changes with the size of the binary. The numbers quoted in the panels indicate the total fraction of $\mu$TDEs integrated over all phases and impact parameters shown.}    
    \label{fig:mu_phase}
\end{figure*}

Our results show that for interactions where the impact parameter is larger than the size of the BBH, the scattering is almost identical to that from a single BH. But for impact parameters smaller than this, we see two different branches of the post-scattered orbits. One branch is that of stars that get scattered towards the SMBH, similar to being scattered by a single BH, but with a wide range of final orbits due to the different phases $\phi$ of the BBH. The other branch is that of stars that get scattered away from the SMBH, which are not present in the single BH scattering.
When the star interacts with the BBH with an impact parameter much smaller than the size of the BBH, the star goes through the BBH, and can get pulled in opposite directions from the individual BHs, one pulling it closer to the SMBH and the other pulling it away. The second branch results from scenarios where the star is closer to the BH pulling the star away from the SMBH, and the net effect is a post-scattered orbit with a larger pericenter distance from the SMBH.

Next, with a similar analysis to that of the standard TDE scenario, we use our scattering experiments to generate phase-space diagrams in order to identify the regions in the parameter space leading to $\mu$TDEs. In Figure \ref{fig:mu_phase}, the left column shows the phase-space leading to $\mu$TDEs (blue dots) for the BBH size $a_{\rm BBH}=$ 0.01$R_{\rm H}$, while the right column shows the same for $a_{\rm BBH}=$ 0.1$R_{\rm H}$. Unlike standard TDEs, which are much more likely to occur as a result of the nearly radial orbits of retrograde stars, $\mu$TDEs can sizeably result from a wide range of eccentricities. More importantly, unlike standard TDEs in AGN disks, where prograde scattering barely contributes, $\mu$TDEs occur frequently in both prograde and retrograde scatterings. Therefore, we explore both prograde and retrograde scatterings. We display the results at the representative scattering location $r$ = $10^3r_{g}$.

We assume an initially circular orbit for stars in prograde orbits (initial $e=0$), while for retrograde orbits we consider a typical case of a highly eccentric orbit, with initial $e=0.998$. The top panels of Figure \ref{fig:mu_phase} display our scattering experiment results for retrograde star orbits, while in the bottom panels we show our results for prograde star orbits. For each case, the left and right panels illustrate the dependence of the scattering experiments on the binary size. At the bottom of each plot, we indicate the total fraction of events (i.e. integrated over the full phase space shown) that results in $\mu$TDEs.

For prograde orbits, the relative velocity between the BBH and the star is small. {This leads to a larger gravitational focusing factor as described in Section~\ref{sec:utde}. As a result, $\mu$TDEs appear in a larger region of the parameter space.} 

In the case of retrograde orbits, the star is in a highly eccentric, nearly radial path toward the SMBH, and the relative velocity between the BBH and the star is large. { As a result, the gravitational focusing effect is much weaker than the prograde scatterings. Due to this reason, when the impact parameter is small, the star can easily pass through the BBH without encountering either of the component BHs.}

\subsection{Cross-sections}

The cross-section of a star undergoing a TDEs is given by
\begin{eqnarray}
    \sigma_{\rm TDE} = \int_{\Sigma L_{\rm TDE}} db \label{eq:sigma}
\end{eqnarray}
where the integration is performed over the impact parameter $b$, identifying the regions $\Sigma L_{\rm TDE}$ in the parameter space where  TDEs  are observed. {The impact parameter is expressed in units of the Hill radius, $R_{\rm H}$}, thus the cross-section is calculated in units of $R_{\rm H}$ for standard TDEs, while for $\mu$TDEs the cross-section is calculated in units of $R_{\mu TDE}$ corresponding to a 60$M_{\odot}$ BH (note the units of length - rather than length square, since we are in a coplanar geometry). This integration needs to be performed over the entire range of initial pericenter distances of the orbit of the star, or equivalently over the entire range of initial eccentricities. 

From above, the probability distribution of the orbital eccentricity of stars will depend on whether they are in prograde or retrograde orbits. The former, after a possibly transient period, will eventually settle in circular orbits, and (relatively) slowly migrate through the disk
due to gas torques, subject to occasional perturbations from in-disk binary-single scattering.
The latter (stars on retrograde orbits) will develop a very high eccentricity. For these stars in retrograde orbits it can be assumed that the initial orbital angular momentum has a uniform probability distribution function, such that $p(L)$ = constant
{(see Appendix A for a justification). }
Then the eccentricity  probability distribution $p(e)$ readily follows from 
\begin{eqnarray}
p(e) = p(L) |dL/de |\propto \frac{e}{\sqrt{1-e^2}} \,,
\label{eq:p_ecc}
\end{eqnarray}
where $p(e)$ is normalized to unity using $\int_{0}^{1} de p(e)$ = 1. The cross-section is computed by drawing random values of the eccentricity from this distribution.

%\begin{figure}
%    \includegraphics[width=\columnwidth]{totalsigma_ecc_final.png}
%   \caption{Cross-sections for TDEs as functions of the distance $r$ from the SMBH for different initial eccentricities of the disrupted star. The solid line shows results for a BBH of size $a_{\rm BBH}=$ 0.1$R_{\rm H}$, the dashed line shows results for a BBH of size $a_{\rm BBH}=$ 0.01$R_{\rm H}$, and the dotted line shows results for a single BH.}
%    \label{fig:totalsigma_ecc}
%\end{figure}

%\begin{figure}
%    \includegraphics[width=\columnwidth]{totalsigma_final.png}
    
%    \caption{Cross-sections for the final pericenter distance of the star from the SMBH as functions of the distance $r$ from the SMBH. Initial eccentricities are drawn from a uniform probability in angular momentum for the incoming star. The solid lines show results for a BBH of size $a_{\rm BBH}=$ 0.1$R_{\rm H}$, the dashed lines show results for a BBH of size $a_{\rm BBH}=$ 0.01$R_{\rm H}$, and the dotted lines show results for a single BH, however all three cases look indistinguishable. The blue lines show the cross-section for interactions where the final pericenter distance is less than the tidal disruption radius, thus triggering a TDE.}
%    \label{fig:cs}
%\end{figure}

\begin{figure}
    \includegraphics[width=\columnwidth]{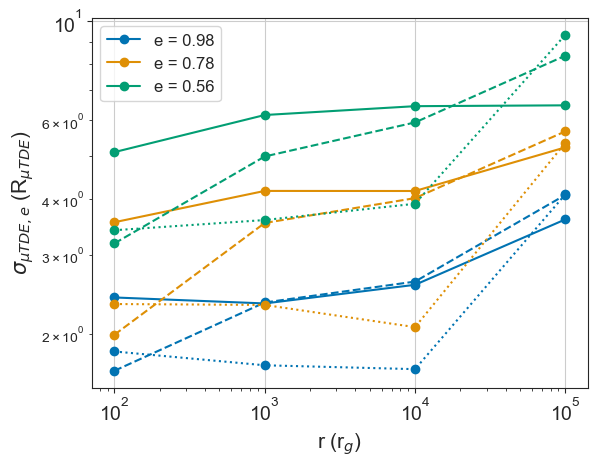}
    \caption{$\mu$TDE (disruption by the stellar mass BH) cross-section as a function of the distance $r$ from the SMBH for different initial eccentricities of the disrupted star  (incoming in a retrograde orbit).
    The solid line shows results for a BBH of size $a_{\rm BBH}=$ 0.1$R_{\rm H}$, the dashed line shows results for a BBH of size $a_{\rm BBH}=$ 0.01$R_{\rm H}$, and the dotted line shows results for a single BH.}
    \label{fig:musigma_ecc}
\end{figure}

For the computation of the cross-section of standard TDEs onto the SMBH, we only focus on cases where the TDEs are induced due to the scattering from the BH/BBH. 
This is determined by scenarios where the pre-scattering pericenter distance of the star is larger than the SMBH tidal disruption radius, but post-scattering it becomes smaller than it, thus leading to a scattering-induced TDE. We compute this cross-section for three different sizes of the scattering BBH: $a_{\rm BBH}$ = 0.1$R_{\rm H}$, $a_{\rm BBH}$ = 0.01$R_{\rm H}$, and $a_{\rm BBH}=0$, which effectively corresponds to the case of a single BH of total mass equal to the sum of the two BHs in the binary. Our numerical experiments show that this cross-section is largely independent of the size of BBH and its distance from the SMBH, and is on the order of $\sigma_{\rm TDE} \sim 1\times 10^{-3}R_{\rm H}$. So, larger mass BBH ($R_{\rm H} \propto M_{\rm BBH}^{1/3}$) at larger radial disk distance ($R_{\rm H} \propto r$) will be more efficient TDE generators per scattering encounter (although more frequent encounters occur at small $r$).

For the computation of the cross-section of $\mu$TDEs, we assume circular orbits for stars in prograde orbits, and the same probability in Eq.~\ref{eq:p_ecc} for stars in retrograde orbits. 
In order to gain physical insight into the range of orbital parameters more likely to yield $\mu$TDEs, we begin by showing the cross-sections at fixed values of the initial eccentricity $e$, for each distance $r$ of the scatterer from the SMBH.

Unlike standard TDEs, which are largely favored by radial orbits, $\mu$TDEs are more easily favored by lower eccentricities. For high eccentricities, the star has a near straight-line path. Due to the large relative velocity between the star and the BBH, the time spent by the star in close vicinity of the BBH is small. As the eccentricity is lowered, the star takes a more curved path. {The relative velocity between the star and the BBH is significantly reduced; as a result, the star} spends more time close to the BBH, leading to a higher probability of being disrupted by either of the component BHs. This effect is even stronger when the BBH size is large since a larger BBH allows the star larger window of time to encounter a component BH. If the BBH is replaced by a single BH, this window of time is smaller and the probability of a $\mu$TDE is smaller. This is illustrated by Figure~\ref{fig:musigma_ecc}, where we show our results for the $\mu$TDE cross-sections for three different initial eccentricities of the star (incoming in a retrograde orbit), and for two different sized BBH and single BH scattering. 

In terms of the trend with distance $r$ of the BH/BBH from the SMBH we note that, generally, the rate of $\mu$TDEs stays nearly constant in inner regions and increases as $r^{1/2}$ in the outer regions, as also found in the two-body case by \citet{Wang_2023}. This is due to the fact that, in the outer region, the orbital velocity difference between the star and the BH/BBH around the SMBH is significantly smaller than in the inner region. Thus, the gravitational focusing effect in the outer region is stronger. Since the star will be disrupted at the fixed radius $R_{\mu \rm TDE}$, a stronger focusing effect in the outer region leads to a larger cross-section of the $\mu$TDE. 

\begin{figure}
    \includegraphics[width=\columnwidth]{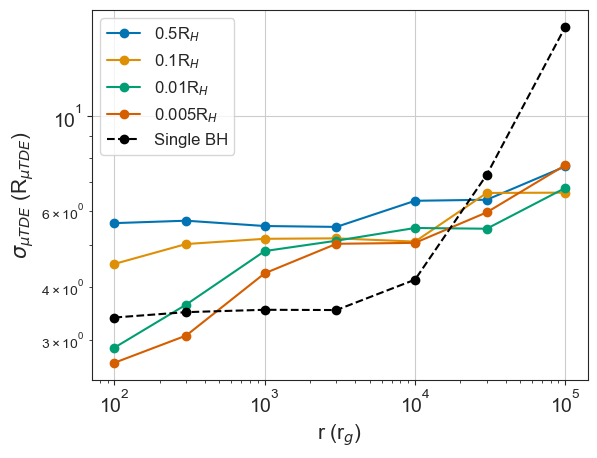}
    \caption{Cross-sections of $\mu$TDEs for retrograde stars as a function of the distance $r$ of the sBH/BBH location from the SMBH. Initial eccentricities are drawn from a uniform probability in angular momentum for the incoming star.}
    \label{fig:csmu}
\end{figure}

The net cross-section, weighted by the probability distribution $p(e)$ at each scattering location $r$, is displayed in Figure~\ref{fig:csmu} for a range of binary sizes. Following the trend in Figure \ref{fig:musigma_ecc}, the magnitude of the net cross-section shows a small dependence on the distance $r$; also, the larger BBH has a slightly larger cross-section.

\begin{figure}
    \includegraphics[width=\columnwidth]{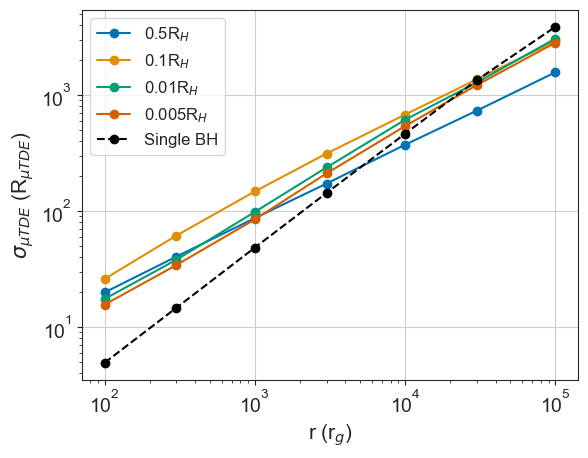}
    \caption{Cross-section of $\mu$TDEs for stars in prograde, circular orbits as functions of the distance $r$ of the sBH/BBH location from the SMBH.}
    \label{fig:csmu_pro}
\end{figure}

Last, Figure \ref{fig:csmu_pro} shows the scattering experiment results for the cross-section of $\mu$TDEs for stars in prograde, circular orbits. Due to the smaller relative velocities, the gravitational focusing effect is especially strong in prograde scatterings, which contribute most of the $\mu$TDEs in AGN disks. 
This effect is even stronger in the outer regions of the disk since the relative velocity between the star and the BBH decreases with increasing distance $r$, and thus the cross-section displays a steep increase with  $r$.

\subsection{From cross-sections to event rates}
We begin by calculating the event rate per system (per BH for single scattering and per BBH for binary scattering) by using:
\begin{eqnarray}
\mathcal{\gamma} = 2H\sigma_{\rm TDE} n_* v
\label{eq:rate}
\end{eqnarray}
where  $n_*$ represents the number density of stars within the disk, and $v$ is the relative velocity between the stars and the BH/BBH center of mass. Since $H/r\ll 1$, the scatterings are almost identical within the scale height $H$ of the disk at a given $r$. We obtain the normally defined cross-section (in unit of area) by multiplying $\sigma_{\rm TDE}$ by the thickness of the disk $2H$, as given by Equation~\ref{eq:H}.

Given the poorly constrained population of stars in AGN disks, we use the number density $n$ of stars in the NSC  to provide a rough approximation to the number density $n_*$ of stars in the AGN disks.

The number density $n$ of stars in the NSC follows a power-law distribution characterized by the index $\gamma_{\rm NSC}$, as given by
\citep{Merritt2004,Merritt2013}:
\begin{eqnarray}
n = \frac{3-\gamma_{\rm NSC}}{2\pi} \frac{M_{\rm SMBH}}{m} r_h^{-3} \Big(\frac{r}{r_h}\Big)^{-\gamma_{\rm NSC}}.
\label{eq:den}
\end{eqnarray}
Here, $m$ represents the average mass of a star in the NSC, and $r_h$ is the gravitational influence radius of the SMBH. 

{For the relative velocity between the star and the BH/BBH ($v$), in retrograde scatterings (i.e., the stars move in the opposite direction to the BH/BBH), we use $2 v_{\rm k}$, where $v_{\rm k}$ is the local Keplerian velocity. For prograde scatterings, we use the shear velocity difference $\frac{d v_{\rm k}}{dr}\langle \Delta r \rangle$, where $\langle \Delta r \rangle \sim (1/n)^{1/3}$ is the average distance between objects in the disk, which can be obtained from Equation~\ref{eq:den}.

To obtain the global rate of the scattering induced TDE/$\mu$TDE, we integrate $\gamma_{\rm TDE}$ over the disk volume
\begin{eqnarray}
{\Gamma}\sim \int n_{\bullet}\gamma dV \sim  \int_0^{r_{\rm max}} n_{\bullet}\gamma 2\pi r 2H(r) dr\,,
\end{eqnarray}
where $n_\bullet$ is the number density of BH/BBH in the AGN disks, and $r_{\rm max}$ is the minimum between $r_h$ and $10^5 r_g$, which is the maximum radius of our scattering experiments, set as a conservative limit given the uncertain exact value of the disk outer radius. For simplicity (and lack of detailed estimates in the literature), we assume that stars and BHs/BBHs follow the same density profile of the stars. We note however that this is most likely not strictly the case, since, due to mass segregation, BHs tend to migrate towards the inner cluster regions, where their abundance compared to that of the stars thus becomes relatively higher compared to the outer regions \citep{Generozov2018}. To keep transparent the dependence of our rates $\Gamma$ on $n_\bullet$, we define the  BH-to-star ratio $f_{\bullet}=n_\bullet/n$, and express the rates in terms of this quantity. We will use as a reference in the rescaling the value $f_{\bullet}=0.01$, comparable to the fraction found in the inner NSC regions by \citet{Generozov2018}.

For standard TDEs by the SMBH, which we studied for retrograde stars, we estimate about $(6\times 10^{-5}$ - $5\times 10^{-2})(f_\bullet/0.01)$ scattering induced TDEs per year, largely independent of whether the scatterer is a single BH or a BBH, and also largely independent of the size of the BBH. The range of rates corresponds to the interval $\gamma_{\rm NSC}\sim 1-2$, respectively. 
We note (not surprisingly) that this scattering-induced TDE rate, while being relatively high during this AGN 'turn-on' phase, when highly eccentric stars are produced, it is however subdominant with respect to the main TDE flow due to stars on nearly radial orbits which migrate towards the SMBH without help from scattering, and which has been discussed in detail in previous works \citep{McKernan2022,Wang_2023}.

For the rate of $\mu$TDEs and encounters with retrograde stars, 
we estimate about $(2\times 10^{-5}$ - $3\times 10^{-2})(f_\bullet/0.01)$ events per year if the scatterer is a binary, while for single BH scattering the rate increases to $(5\times 10^{-5}$ - $4\times 10^{-2})(f_\bullet/0.01)$ per year. For prograde stars, we estimate $(1\times 10^{-4}$ - $4\times 10^{-2})(f_\bullet/0.01)$ $\mu$TDEs events per year again largely independent of whether the scatterer is a single BH or a BBH. Likewise above, the interval for the rates corresponds to the range $\gamma_{\rm NSC}\sim 1-2$. 

The reported rate upper limits mentioned above apply to a $10^8 M_\odot$ SMBH, which is a representative SMBH for AGNs. For SMBHs of other masses, the rates can be obtained using the following scalings:
\begin{eqnarray}
\dot{M} &\propto& M_{\rm SMBH}\\
H &\propto& \Omega_d^{-1/3} \propto M_{\rm SMBH}^{-1/6}\\
r_h &\propto&  M_{\rm SMBH} \sigma_{\rm NSC}^{-2} \propto M_{\rm SMBH}^{0.54}\\
n &\propto& (3 - \gamma_{\rm NSC}) M_{\rm SMBH}^{0.54(\gamma_{\rm NSC}-3)+1} \\
v_{\rm ret} &\propto& v_k \propto  M_{\rm SMBH}^{1/2}\\
v_{\rm pro} &\propto& v_k n^{-1/3} \propto M_{\rm SMBH}^{1/6-0.18(\gamma_{\rm NSC}-3) }\\
\sigma_{\rm TDE} &\propto& R_{\rm TDE} \propto  M_{\rm SMBH}^{1/3}\\
\sigma_{\rm \mu TDE} &\propto&  M_{\rm SMBH}^{0}\,.
\end{eqnarray}
}
For $\sigma_{\rm NSC}$, we use the M-$\sigma$ relation $\sigma_{\rm NSC}\propto M_{\rm SMBH}^{1/4.38}$\citep{Kormendy2013}.

For prograde scatterings (denoted (p,TDE)), the rates scale according to the following relationships (using $r_{\rm max}=r_h$ for generality):
\begin{eqnarray}
\gamma_{\rm p,TDE} &\propto&  M_{\rm SMBH}^{0.36(\gamma_{\rm NSC}-3)+4/3}\\
\gamma_{\rm p, \mu TDE} &\propto&  M_{\rm SMBH}^{0.36(\gamma_{\rm NSC}-3)+1}\\
\Gamma_{\rm p,TDE} &\propto&  M_{\rm SMBH}^{1.9}\\
\Gamma_{\rm p, \mu TDE} &\propto&  M_{\rm SMBH}^{1.02}\,.
\end{eqnarray}

For retrograde scatterings (denoted (r,TDE)), the rates scale as follows:
\begin{eqnarray}
\gamma_{\rm r,TDE} &\propto&  M_{\rm SMBH}^{0.54(\gamma_{\rm NSC}-3)+5/3}\\
\gamma_{\rm r, \mu TDE} &\propto&  M_{\rm SMBH}^{0.54(\gamma_{\rm NSC}-3)+4/3}\\
\Gamma_{\rm r,TDE} &\propto& 
M_{\rm SMBH}^{2.23}\\
\Gamma_{\rm r,\mu TDE} &\propto& 
M_{\rm SMBH}^{1.36}  \,.
\end{eqnarray}

These scalings provide a framework for estimating the rates of TDEs and $\mu$TDEs for SMBHs of various masses based on the dynamics within the AGN disks.

\section{Discussion}
Using idealized three body scattering experiments, we have explored a  wide range of scenarios that can generate an AGN TDE around the central SMBH, or an AGN $\mu$TDE from stellar-mass BHs embedded in an AGN disk. We focused on the scattering of retrogade orbiting stars by BH binaries (BBHs), given their large {geometric} cross-sections, and the likelihood of BBHs in an AGN disk. We investigated a range of BBH sizes and BBH locations in the disk.

Standard loss-cone filling by dynamical encounters is inefficient in AGN. However, stars on retrograde orbits should be about half the initial stellar population early ($\leq 0.1$~Myr) in the AGN lifetime. Fully embedded retrograde stars should rapidly acquire a very high eccentricity and can either end up directly disrupted by the SMBH, or be scattered into the AGN loss-cone \citep{McKernan2022}. Thus, AGN TDEs should overwhelmingly arise early in an AGN lifetime before the retrograde orbiters disappear. 
We find TDE rates of $\sim (6\times 10^{-5} - 5\times 10^{-2})(f_\bullet/0.01)$ ${\rm AGN^{-1} \rm{yr}^{-1}}$ which spans comparable to or significantly higher than the 'naked' TDE rate O($10^{-4}{\rm galaxy }^{-1}{\rm yr}^{-1}$). The magnitude of the AGN TDE rate is determined by a combination of the slope of the NSC density profile and the outer disk radius, with very little dependence on the BBH size.  AGN TDEs from retrograde orbiters may display significantly higher luminosities than naked TDEs, or prograde TDEs, with optical/UV luminosity
up to $\sim 10^{44}$~erg~s$^{-1}$ and should significantly heat the inner AGN disk \citep{McKernan2022}, which may be detectable in large AGN surveys. An additional source of AGN TDEs may be due to stars on prograde orbits, via multiple scatterings, which is deferred to future work.

The dynamical conditions leading to $\mu$TDEs are both qualitatively
and quantitatively different than for standard TDEs. In this case,
lower eccentricities are generally favored, and hence there can be
contributions also from prograde orbiters. We find $\mu$TDEs rates of $\sim (1\times 10^{-4} - 4\times 10^{-2})(f_\bullet/0.01)$ 
${\rm AGN}^{-1}$~yr$^{-1}$ (prograde), and  $\sim (2\times 10^{-5} - 3\times 10^{-2})(f_\bullet/0.01)$ ${\rm AGN}^{-1}$~yr$^{-1}$ (retrograde). While AGN TDEs should be dominated by stars on retrograde orbits and appear mostly early in the AGN lifetime, $\mu$TDEs from stars on prograde orbits are present throughout the AGN lifetime, thus providing a great probe to study the population of embedded objects over the entire AGN cycle.
Also importantly, while retrograde encounters tend to be hyperbolic and hence not expected to lead to  flaring since the debris get unbound (see e.g. \citealt{Hayasaki2018}), prograde encounters are eccentric, and the debris will thus remain bound, hence allowing the needed rapid accretion to power TDE flares. Recent work suggests that debris, particularly from retrograde TDEs, can be fully mixed into dense AGN disk, and the perturbation of the inner disk by the TDE will generate a state-change, likely accounting for some observations of changing look AGN \citep{Taeho24}.

%$\mu$TDEs from prograde encounters are therefore a great probe to study the population of embedded objects over the entire AGN lifetime.

The  duration, light curves and spectra of TDEs and $\mu$TDEs are expected to differ in multiple ways. The lower BH mass in $\mu$TDEs results in accretion rates which can exceed the Eddington threshold by up to
a factor of $10^{5}$ or more \citep{Metzger2016,Kremer2019,Lopez2019,
Wang2021TDE,Kremer2022}. The combination of high accretion rates and
low BH masses in $\mu$TDEs results in hotter disks compared to those
of standard TDEs \citep{Wang2021TDE}, leading to a high X-ray
luminosity. Most importantly, the hyper-Eddington accretion rates are
likely to create conditions conducive to the launch of relativistic
outflows which may then dissipate yielding significant $\gamma$ ray
and X-ray emission \citep{Murase2016}. Such emission, longer but
weaker than that of the standard GRBs, may rather resemble that
of the ultralong GRBs \citep{Perets2016}.

Finally, detection from AGN disks eventually requires the radiation to
pass through the dense disk material on its way to the observer.
Depending on the AGN disk mass and scatterer distance from the SMBH,
the radiation can emerge unaltered, marginally diffused, or highly
diffused \citep{Perna2021,Zhu2021,Wang2022GRB}, resulting in increasingly
dimmer but longer-lived transients. Future searches for signatures of AGN TDEs and $\mu$TDEs in large samples of AGN will help constrain the dynamics of embedded populations in AGN disks over typical AGN lifetimes.

\section*{Acknowledgements:}
CP and RP acknowledge support by NSF award AST-2006839. 
BM \& KESF are supported by NSF AST-2206096 and NSF AST-1831415 and Simons Foundation Grant 533845 as well as Simons Foundation sabbatical support. The Flatiron Institute is supported by the Simons Foundation. YW is supported by the Nevada Center for Astrophysics. We thank the anonymous referee for an exceptionally
constructive and thoughtful report which helped to considerably improve the paper.

\section*{Data Availability}
Any data used in this analysis are available on reasonable request from the first author (CP).


\appendix

{

\section{Initial orbital angular distribution of retrograde stars}

The distribution of angular momentum can be parameterized by a power-law index, denoted as $\alpha$, where $p(L)\propto L^\alpha$. This parameterization leads to an eccentricity distribution described by $p(e)\propto e({1-e^2})^{(\alpha-1)/2}$. When $\alpha$ equals 1, known as the thermal distribution, the eccentricity distribution simplifies  to $p(e)\propto e$. For values of $\alpha$ less than 1, the distribution becomes superthermal, indicating that the eccentricity is predominantly high. It can be seen that for any value of $\alpha$ less than or equal to -1, the distribution function $p(e)$ becomes divergent, with the integral of $p(e)$ from 0 to 1 approaching infinity. This suggests that the distribution effectively becomes a delta-like function, specifically $\delta(e=1)$.

Notably, $\alpha=0$, which corresponds to a uniform distribution of angular momentum, is the last superthermal distribution with an integrable integer value of $\alpha$ over the range of $e$ from 0 to 1. Since retrograde orbits in the fast inspiral regime will inevitably enter the tidal radius, akin to $\delta(e=1)$, any integer value of $\alpha$ less than 0 results in the same TDE  rate. The $\alpha=0$ distribution is the most convenient for sampling and provides a lower limit for the TDE rate in this regime. Therefore, we selected this distribution. However, it is important to acknowledge that determining the true value of $\alpha$ for the fast inspiral regime requires extensive hydrodynamic simulations and is sensitive to the specific accretion disk model used. With that in mind,
and since any value of $\alpha$  less than or equal to 1 yields the same effective TDE rate, adopting $\alpha=0$ in this paper, which yields a lower limit to the TDE rate, should be reasonable.

}

\bibliographystyle{mnras}
\bibliography{ref}

%%%%%%%%%%%%%%%%%%%%%%%%%%%%%%%%%%%%%%%%%%%%%%%%%%

%%%%%%%%%%%%%%%%% APPENDICES %%%%%%%%%%%%%%%%%%%%%

%%%%%%%%%%%%%%%%%%%%%%%%%%%%%%%%%%%%%%%%%%%%%%%%%%

% Don't change these lines
\bsp	% typesetting comment
\label{lastpage}
\end{document}